\documentclass[twocolumn,showpacs,preprintnumbers,amsmath,amssymb]{revtex4}
\usepackage{amsmath,amssymb,graphics,epsfig,subfigure}
\usepackage{color}

\begin{document}

\thispagestyle{empty}

\begin{center}

\title{Gravity/thermodynamics correspondence via black hole shadows}

\date{\today}
\author{Shao-Wen Wei$^{a,b}$ \footnote{E-mail: weishw@lzu.edu.cn},
Yu-Xiao Liu$^{a,b}$ \footnote{E-mail: liuyx@lzu.edu.cn}}

\affiliation{$^{a}$Lanzhou Center for Theoretical Physics, Key Laboratory of Theoretical Physics of Gansu Province,
Key Laboratory of Quantum Theory and Applications of MoE,
Gansu Provincial Research Center for Basic Disciplines of Quantum Physics, Lanzhou University, Lanzhou 730000, China,\\
$^{b}$Institute of Theoretical Physics $\&$ Research Center of Gravitation,
School of Physical Science and Technology, Lanzhou University, Lanzhou 730000, China}

\begin{abstract}
The shadow of a black hole serves as a pristine window into the strong-gravity regime, with cuspy feature emerging as a smoking-gun signature of physics beyond the Kerr paradigm. In this paper, we extend the work of [arXiv:2601.15612 [gr-qc]] and study the detailed properties of the cuspy shadow by using the parametric expressions of the shadow boundary. From a topological perspective, we provide a rigorous topological classification of these shadows, categorizing them into distinct ``rectangular" and ``8-shape" topologies. Crucially, we establish a formal gravity/thermodynamics correspondence by mapping the cuspy shadow to the swallowtail behavior observed in thermodynamic free energy. We demonstrate that the self-intersection of the shadow boundary, marking a geometric phase transition, can be precisely determined through three independent but equivalently thermodynamic-like approaches. Furthermore, we analytically derive the critical exponents governing the emergence of these cusps, revealing that they are consistent with the mean-field universality class. Our results suggest that the observational features of black hole shadows are deeply rooted in the underlying gravitational thermodynamics, offering a novel framework to probe the fundamental nature of spacetime.
\end{abstract}


\maketitle
\end{center}

\section{Introduction}
\label{secIntroduction}

The black hole shadow represents a hallmark signature of the strong-field gravitational regime, arising from the extreme deflection and capture of null geodesics in the vicinity of the event horizon \cite{Synge,Luminet}. As a robust observation, the shadow's shape, specifically its characteristic size and geometric deformations, encodes important information regarding the underlying spacetime metric. Consequently, the analysis of shadow features provides a powerful test to constrain the black hole parameters, including mass, spin, and potential exotic charges \cite{Bardeen,Chandrasekhar,Hioki,Amarilla,Nedkova,Atamurotov,Wei,Cunhaz,Ghosh,Chen,Guo,Tsukamoto,Battista,GaoHu}. For a comprehensive review of the theoretical foundations and recent developments in shadow physics, we refer the reader to Refs. \cite{CunhaHerdeiroa,Perlick,JingQian}.

These theoretical foundations have gained remarkable significance with the recent observational breakthroughs by the Event Horizon Telescope (EHT). The images of $M87^*$ and Sgr $A^*$ have provided unprecedented confirmation of general relativity in the strong-field regime, allowing for constraints on fundamental parameters such as black hole mass, spin, and charge \cite{EHT,EHTb,EHTc,EHTd}. These milestones have effectively stimulated the theoretical shadow modeling. Recent studies have moved beyond vacuum cases to incorporate the complex influences of accretion disks, astrophysical material flows, and dark matter halos \cite{Gralla,Freese,Zinhailo,SenGupta,Wangb,Zhangb,Junior,Hou,Kuangb,Yangb,CaoLi}. These developments provide the necessary context for exploring more exotic shadow features that could signal deviations from standard general relativity.

In a static, spherically symmetric background such as the Schwarzschild spacetime, the black hole shadow appears to a distant observer as a geometrically perfect circle. The introduction of black hole spin, however, fundamentally alters this picture. The frame-dragging effect induces a pronounced asymmetry in the trajectories of light rays, distorting the shadow into a characteristic ``D-shape" \cite{Chandrasekhar}. Despite this, within the standard Kerr paradigm, such morphological deviations remain relatively subtle across much of the parameter space, becoming observationally significant only as the black hole spin approaches its extremal limit.

For a spherically symmetric black hole, the shadow boundary is uniquely determined by the photon sphere. Due to the high degree of spacetime isometry, every light ray constituting this boundary follows a circular orbit at a fixed radius, which can be effectively mapped onto an equatorial trajectory. In contrast, a spinning black hole possesses only axial symmetry, significantly enriching the underlying null geodesic flow. As established in Ref. \cite{CunhaHerdeiroRadu}, such backgrounds admit a more general class of fundamental photon orbits, bounded null geodesics, that do not necessarily lie in the equatorial plane. These fundamental photon orbits, which dictate the shadow's shape, are typically spherical orbits with a constant radial coordinate. However, this radius no longer remains the same and instead varies continuously with the orbit's inclination angle or Carter constant. In this framework, the light rings represent the specific subset of these spherical orbits that are confined to the equatorial plane.

Reference \cite{CunhaBerti} was the first to consider the topological charge, winding number, for the light ring in the background of a compact object, and then the study was generalized to the black holes \cite{Cunha}. The results confirmed that, within any stationary and axisymmetric Kerr-like background, there must exist at least one unstable light ring outside the event horizon for each sense of rotation, prograde and retrograde. This existence theorem has been successfully generalized to non-rotating black holes in asymptotically flat, AdS, and dS spacetimes \cite{Weisw}, emphasizing the universal nature of light rings/photon spheres as fundamental features of the strong-gravity environment.

Significantly, while the spherical orbits mentioned above are inherently radially unstable, a profound transition occurs in various non-Kerr scenarios. In the presence of Proca hair, dark matter halos, varying gravitational constants, or wormhole geometries \cite{CunhaHerdeiroRadu,WangChen,GyulchevYazadjiev,Qian,Ohta,ChengZhao,ChengYang}, stable spherical orbits can emerge. The existence of these stable spherical orbits fundamentally alters the shadow's morphology, leading to the formation of ``cuspy" shadows. For observers at a finite distance, these features can exhibit as intricate ``eyebrow" structures or even fragmented secondary shadows. Notably, the gravitational lensing effect is dramatically amplified in regions where stable orbits dominate \cite{CunhaHerdeiroRadu}. Such distinct topological anomalies suggest that the stable spherical orbits are the definitive factors of cuspy shadows, rendering these images a stringent test for the Kerr hypothesis: any detected deviation would constitute unequivocal evidence for physics beyond general relativity.

In Ref. \cite{Weia}, we investigate the topological properties and self-intersection behavior of the cuspy shadow. While the topological charge of a standard D-shaped shadow remains invariant at $w=1$ independent of the black hole parameters, we demonstrate that the emergence of cusps triggers a transition of this charge to $w=-1$. This shift arises because each of the two cuspy structures effectively acts as a genus, contributing a value of $-1$ to the total topological charge (yielding $w = 1 - 1 - 1 = -1$). To characterize the self-intersection behavior, we propose a ``gravitational equal-area law" formulated by analogy with thermodynamic Maxwell constructions. We also identify a universal critical exponent of $1/2$ near the threshold where the cuspy behavior either emerges or vanishes.

Furthermore, we clarify this gravity/thermodynamics correspondence by taking the spinning Konoplya-Zhidenko (KZ) black hole as a concrete example. By mapping the shadow's cuspy geometry to the swallowtail behavior of thermodynamic free energy, we identify the corresponding gravitational counterparts for temperature, pressure, entropy, and volume (see Table \ref{tab1} for details). Building upon this correspondence, we propose three distinct yet equivalent methods to precisely determine the self-intersection point. Additionally, the critical exponent is analytically derived, further revealing the underlying universality of this geometric phase transition.

This manuscript is organized as follows. In Sec. \ref{ngs}, we derive the null geodesics and present the shadow shapes for the KZ black hole, highlighting the emergence of cuspy structures. In Sec. \ref{Topology}, we examine the topological properties of the shadow, categorizing them into ``rectangular" or ``8-shape" topologies corresponding to topological charges of $1$ and $-1$, respectively. Through this topological framework, the formation mechanism of these cusps is systematically elucidated. The gravity/thermodynamics correspondence is formally established in Sec. \ref{analogy} using thermodynamic-like differential laws. Based on this mapping, the self-intersection points and critical phenomena are explored in Sec. \ref{point} and Sec. \ref{phenomena}. Finally, Sec. \ref{Conclusion} is devoted to concluding remarks and discussions.

\section{Null geodesic and shadow of Konoplya-Zhidenko black holes}
\label{ngs}

Here, we focus on the spinning KZ black holes, which are known to exhibit cuspy shadows \cite{WangChen}. This distinct feature allows for the construction of a correspondence between the black hole's gravity and its thermodynamics.

\begin{figure}
\includegraphics[width=6cm]{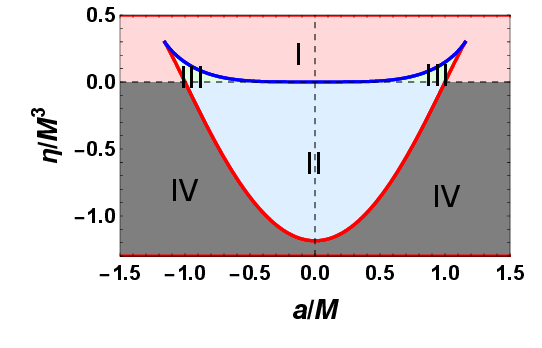}
\caption{Parameter regions corresponding to the number of horizons. In the regions I, II, III, and IV, the KZ black holes have one, two, three, and none horizon(s).}\label{PHSTRPT}
\end{figure}

In the Boyer-Lindquist coordinates, the black hole solution reads \cite{KonoplyaZhidenko}
\begin{eqnarray}
\label{xy}
ds^{2}=& -&\bigg(1-\frac{2Mr^2+\eta}{r\rho^{2}}\bigg)dt^{2}
+\frac{\rho^{2}}{\Delta}dr^{2}+\rho^{2} d\theta^{2}\nonumber\\
&+&\sin^{2}\theta\bigg[r^{2}+a^{2}
+\frac{(2Mr^{2}+\eta)a^2\sin^{2}\theta}{r\rho^{2}}\bigg]d\phi^{2}\nonumber\\
&&-\frac{2(2Mr^2+\eta)a\sin^{2}\theta}{r\rho^{2}}dtd\phi,
\end{eqnarray}
where the metric functions are
\begin{equation}
\Delta=a^{2}+r^{2}-2Mr-\frac{\eta}{r},\quad \rho^{2}=r^{2}+a^{2}\cos^{2}\theta.
\end{equation}
The parameters $M$, $a$, and $\eta$ represent the mass, spin, and deformation parameter of the black hole, respectively. The deviations of the KZ black hole from the Kerr solution are characterized by the parameter $\eta$. Setting $\eta=0$ yields the Kerr black hole. The condition $a\leq M$ must be satisfied for the existence of a horizon; however, for non-vanishing $\eta$, the dimensionless spin $a/M$ may exceed unity. By solving $\Delta=0$, we can obtain the radii and number of horizons. In Fig. \ref{PHSTRPT}, we illustrate the parameter regions corresponding to the number of horizons. Specifically, regions I, II, and III exhibit one, two, and three horizon(s), respectively, while region IV does not support the existence of a horizon. Notably, for positive $\eta$, horizons can exist for arbitrarily large black hole spins. The blue and red curves are given by
\begin{equation}
 \eta_\pm=\frac{2}{27}\left(9a^2M-8M^3\pm\sqrt{(4M^2-3a^2)^3}\right).
\end{equation}
The bottom of the red curve is located at (0, $-32M^3/27$), while these two cusps are found at ($\pm2M/\sqrt{3}$, $8M^3/27$).

The null geodesic of the spinning KZ black holes is separated. The Hamiltonian of the photons propagating in this background is
\begin{eqnarray}
 H=\frac{1}{2}g^{\mu\nu}p_{\mu}p_{\nu}=0.\label{hami}
\end{eqnarray}
The momentum is given by $p_{\mu}=g_{\mu\nu}\dot{x}^{\nu}$, where the dot denotes the derivative with respect to the affine parameter. Given the presence of two Killing vectors, $\partial_t$, $\partial_\phi$, we can derive two constants, the energy $E$ and the angular momentum $l$:
\begin{eqnarray}
 p_t=-E=g_{tt}\dot{t}+g_{t\phi}\dot{\phi}, \quad
 p_{\phi}=l=g_{\phi\phi}\dot{\phi}+g_{t\phi}\dot{t}.
\end{eqnarray}
Solving them, we have
\begin{eqnarray}
\label{tfc}
\dot{t}&=&E+\frac{(a^{2}E-al+Er^{2})(2Mr^{2}+\eta)}{\Delta\rho^{2}r},\\
\label{jfc}
\dot{\phi}&=&\frac{aE\sin^{2}\theta(2Mr^{2}+\eta)
+a^{2}lr\cos^{2}\theta}{\Delta\rho^{2}r\sin^{2}\theta}\nonumber\\
&&-\frac{l(2Mr^{2}
-r^{3}+\eta)}{\Delta\rho^{2}r\sin^{2}\theta}.
\end{eqnarray}
Substituting the above solution into (\ref{hami}) and performing a variable separation, we obtain
\begin{eqnarray}
\label{rfc}
\rho^{4}\dot{r}^{2}&=&-\Delta\left(\mathcal{Q}+(aE-l)^{2}\right)+\left(al-(r^{2}+a^{2})E\right)^{2},
\\
\label{thfc}
\rho^{4}\dot{\theta}^{2}&=&\mathcal{Q}-\cos^{2}\theta
\bigg(\frac{l^{2}}{\sin^{2}\theta}-a^{2}E^{2}\bigg),
\end{eqnarray}
where $\mathcal{Q}$ is the separable variable constant known as the Carter constant. It is important to note that Eq. (\ref{hami}) is not always separable in arbitrary spacetime backgrounds. In this context, the Carter constant is associated with the hidden symmetry of the spacetime, which is generated by a second-order Killing tensor $K^{\mu\nu}$ \cite{WangChen}.

The radial motion (\ref{rfc}) can be expressed as
\begin{equation}
 \rho^{4}\dot{r}^{2}+V_{eff}=0
\end{equation}
with the effective potential given by
\begin{equation}
 V_{eff}=\Delta\left(\mathcal{Q}+(aE-l)^{2}\right)-\left(al-(r^{2}+a^{2})E\right)^{2}.
\end{equation}
The boundary of the shadow is determined by the spherical orbits, which requires
\begin{eqnarray}
 V_{eff}=0,\quad V'_{eff}=0, \label{veffr}
\end{eqnarray}
where the prime denotes differentiation with respect to $r$. The stable and unstable orbits correspond to $V''_{eff}>0$ and $V''_{eff}<0$, respectively. By solving these two conditions from (\ref{veffr}), we obtain the reduced angular momentum $\xi=l/E$ and the Carter constant $\sigma=\mathcal{Q}/E^2$:
\begin{eqnarray}
\label{pj}
\xi=\frac{2a^{2}Mr^{2}-a^{2}\eta+2\Delta r^{3}-2Mr^{4}-3\eta r^{2}}{a(2Mr^{2}-2r^{3}-\eta)},\\
\label{q}
\sigma=\frac{r^{4}\left(8a^{2}(2Mr^{3}+3\eta r)-(6Mr^{2}-2r^{3}+5\eta)^{2}\right)}{a^{2}(2r^{3}-2Mr^{2}+\eta)^{2}}.
\end{eqnarray}
Note that $r$ here refers to the radius of the spherical orbits rather than the radial coordinate. Therefore, $r$ is closely dependent on the black hole parameters $M$, $a$, and $\eta$, as well as the photon parameters $\xi$ and $\sigma$.

Considering an observer located far from the black hole at $\theta=\theta_0$, the boundary of the shadow is described by the following celestial coordinates:
\begin{eqnarray}
 \alpha&=&\lim_{r\rightarrow \infty}
   \bigg(-r^{2}\sin\theta_{0}\frac{d\phi}{dr}
      \bigg|_{\theta\rightarrow \theta_{0}}\bigg)
     =-\xi\csc\theta_{0},\label{alpha}\\
 \beta&=&\lim_{r\rightarrow \infty}
   \bigg(r^{2}\frac{d\theta}{dr}\bigg|_{\theta\rightarrow \theta_{0}}\bigg)
     =\pm\sqrt{\sigma+a^{2}\cos^{2}\theta_{0}-\xi^{2}\cot^{2}\theta_{0}}.\nonumber\\\label{beta}
\end{eqnarray}
If the observer locates in the equatorial plane, it reduces to
\begin{eqnarray}
\label{xd1a}
\alpha&=&-\xi,\\
\beta&=&\pm\sqrt{\sigma}.\label{xd1b}
\end{eqnarray}
Therefore, it gives the parametric expressions of the shadow boundary,
\begin{eqnarray}
 \alpha&=&\alpha(r, M, a, \eta),\label{aaa}\\
  \beta&=&\beta(r, M, a, \eta).\label{bbb}
\end{eqnarray}
Examining such parametric expressions, we can obtain the characteristic behaviors of the shadows. For example, the perimeter and area are
\begin{eqnarray}
 \lambda&=&2\int_{r_a}^{r_b}\sqrt{(\partial_r\alpha)^2+(\partial_r\beta)^2}dr,\\
 \mathcal{A}&=&2\int_{r_a}^{r_b}\beta(\partial_r\alpha)dr,
\end{eqnarray}
where $r_a$ and $r_b$ represent the radii of the light rings. The factor ``2" comes from the $\mathcal{Z}_2$ symmetry of the shadow.

According to the study of Ref. \cite{Cunha}, for each sense of rotation, there exists at least one light ring with radii $r_a$ or $r_b$, which is a subclasses of the spherical orbits confined to the equatorial plane with $\mathcal{Q}$=0 or $\sigma=0$. Thus, for each given black hole, we can determine the radii of the light rings by solving Eq. (\ref{veffr}), corresponding to the maximum and minimum values of $r$. By varying $r$, we can then obtain the boundary curve of the shadow in the celestial plane using Eqs. (\ref{xd1a}) and (\ref{xd1b}).

\begin{figure*}
\subfigure[$a/M=0.9$, $\eta/M^3=-0.1$]{\label{Efa5}\includegraphics[width=4cm]{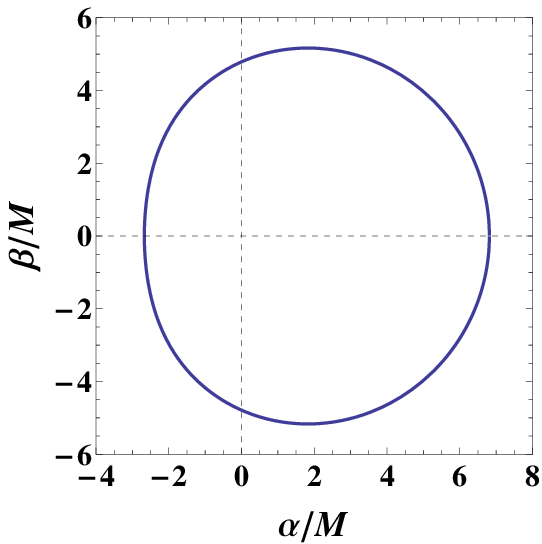}}
\subfigure[$a/M=0.9$, $\eta/M^3=0.08$]{\label{Efb6}\includegraphics[width=4cm]{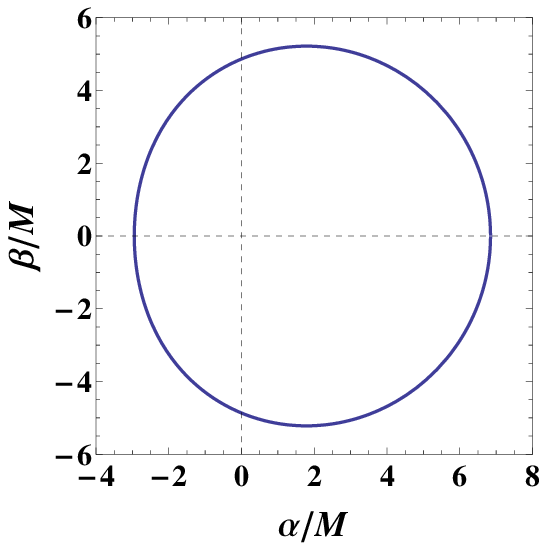}}
\subfigure[$a/M=0.9$, $\eta/M^3=0.2$]{\label{Efc5}\includegraphics[width=4cm]{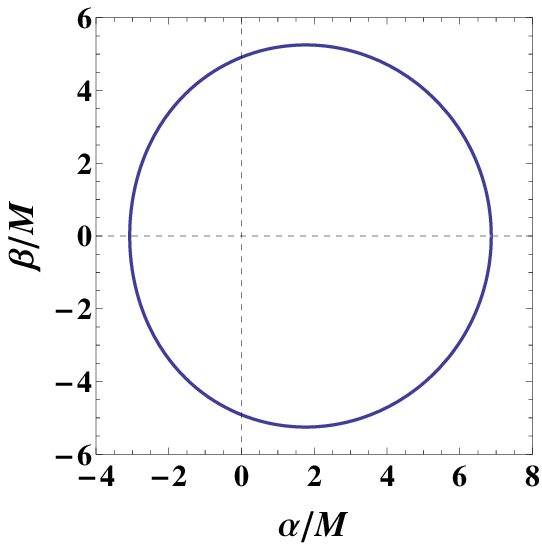}}\\
\subfigure[$a/M=1.1$, $\eta/M^3=0.19$]{\label{Efaa5}\includegraphics[width=4cm]{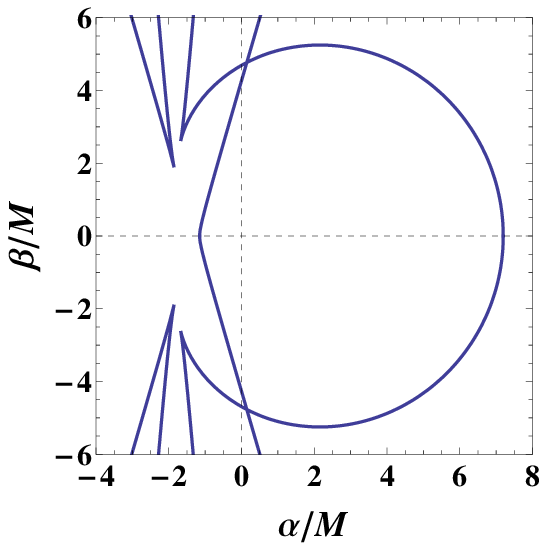}}
\subfigure[$a/M=1.1$, $\eta/M^3=0.22$]{\label{Efba6}\includegraphics[width=4cm]{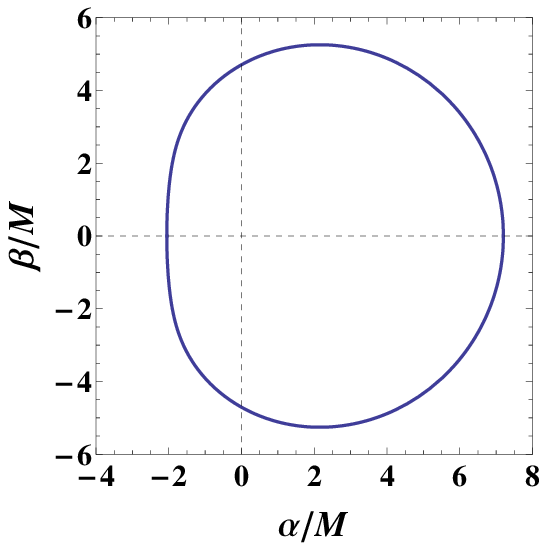}}
\subfigure[$a/M=1.1$, $\eta/M^3=0.25$]{\label{Efac5}\includegraphics[width=4cm]{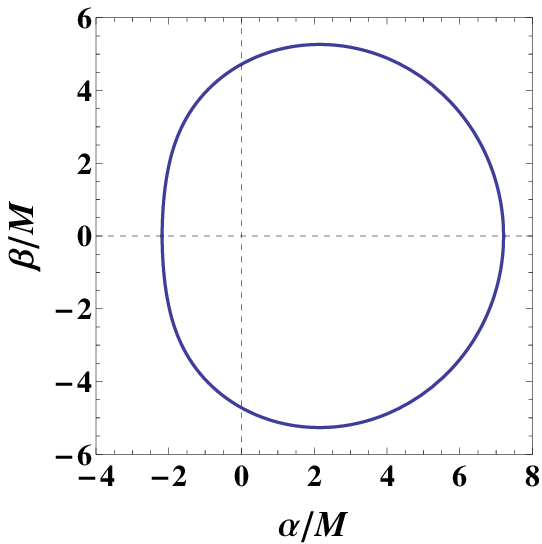}}\\
\subfigure[$a/M=1.2$, $\eta/M^3=0.3$]{\label{Efacd5}\includegraphics[width=4cm]{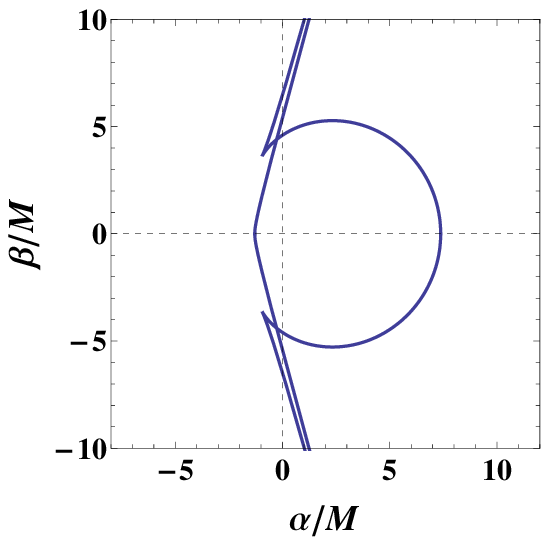}}
\subfigure[$a/M=1.2$, $\eta/M^3=0.33$]{\label{Efbcd6}\includegraphics[width=4cm]{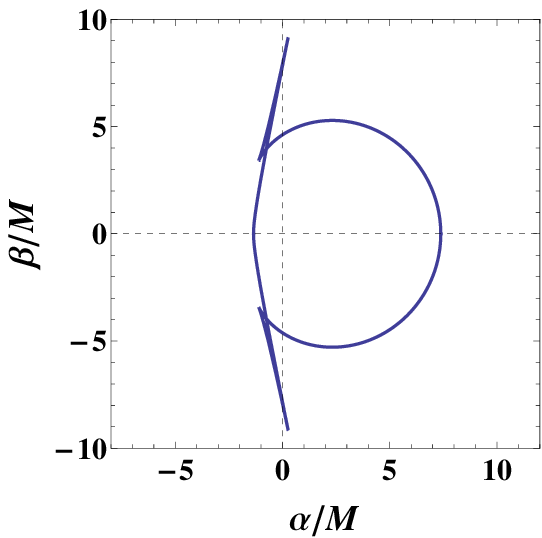}}
\subfigure[$a/M=1.2$, $\eta/M^3=0.4$]{\label{Efccd5}\includegraphics[width=4cm]{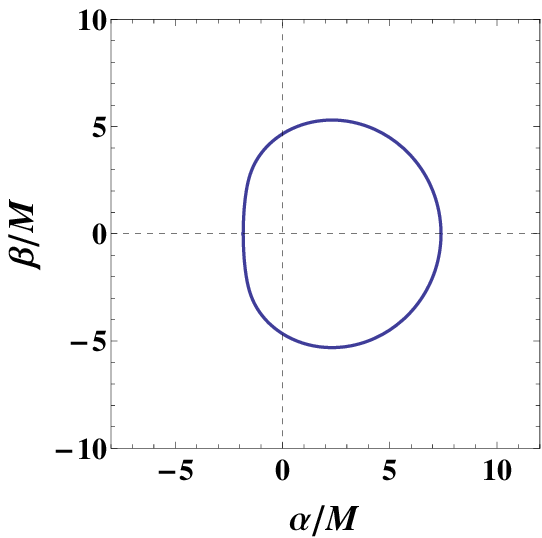}}
\caption{The shadows of spinning KZ black holes with the inclination angle $\theta_0=\pi/2$. In certain cases, the cuspy shadows are exhibited. (a) $a/M=0.9$, $\eta/M^3=-0.1$. (b) $a/M=0.9$, $\eta/M^3=0.08$. (c) $a/M=0.9$, $\eta/M^3=0.2$. (d) $a/M=1.1$, $\eta/M^3=0.19$. (e) $a/M=1.1$, $\eta/M^3=0.22$. (f) $a/M=1.1$, $\eta/M^3=0.25$. (g) $a/M=1.2$, $\eta/M^3=0.3$. (h) $a/M=1.2$, $\eta/M^3=0.33$. (i) $a/M=1.2$, $\eta/M^3=0.4$.}\label{pEfb6}
\end{figure*}

By selecting a few representative values, we illustrate the shadow boundary curves in Fig. \ref{pEfb6}. In Figs. \ref{Efa5}-\ref{Efc5}, we present the results for KZ black holes with $a/M=0.9$, which is less than the maximum spin of Kerr black holes. In this case, the deformation parameter $\eta/M^3$ can take on both positive and negative values. However, the shapes of the shadows remain similar to those of Kerr black holes, exhibiting a D-shape. Notably, for $\eta/M^3=$-0.1, 0.08, and 0.2, the black holes possess two, three, and one horizon(s), respectively; thus, the number of horizons does not influence the appearance of the shadow.

For the Kerr black hole, the maximum spin is $a/M=1$, corresponding to a minimal radius of the outer horizon $r_{h}=M$. In contrast, the KZ black hole can exceed the Kerr limit, with its horizon radius potentially falling significantly below $r_{h}=M$. To clearly illustrate the shadow in this case, we take $a/M$=1.1 and present the shadow boundary shapes in Figs. \ref{Efaa5}-\ref{Efac5} for $\eta/M^3=$0.19, 0.22, and 0.25. For these parameters, the black hole has one, three, and one horizon(s), respectively. Specifically, for $\eta/M^3=$0.19, we have a small black hole with $r_{h}\approx0.24M$. This indirectly leads to the divergences in $\alpha$ and $\beta$ (see Eqs. (\ref{pj}) and (\ref{q})), resulting in a complex shape observable in Fig. \ref{Efaa5}. In the case of the Kerr black hole with $\eta/M^3=0$, the divergent term simplifies to $1/(r-M)$. Since the radius of the spherical orbit is larger than that of the horizon, there is no divergent term, which explains the absence of the complex shape seen in Fig. \ref{Efaa5}. This phenomenon is unique to KZ black holes. For the other two cases with $\eta/M^3=$0.22 and 0.25, the shadow shapes remain similar, regardless of the number of black hole horizons. Further increasing the black hole spin to $a/M=1.2>2/\sqrt{3}$, the KZ black hole is at the right of the cusps, resulting in only one horizon. We discuss the shadows for $\eta/M^3$=0.3, 0.33, and 0.4 in Figs. \ref{Efacd5}-\ref{Efccd5}. For smaller $\eta$, the shadows present cuspy features, while these cuspy behaviors diminish with increasing $\eta$. This occurrence is primarily attributed to the presence of the stable spherical orbits, which we will examine further.

From the above discussion, we note that the cuspy shadow shape is a distinguishing feature of KZ black holes, setting them apart from Kerr black holes. This shape serves as an indicator of the existence of the stable spherical orbits. Therefore, let us turn our attention to examining the stability of these spherical orbits. By applying Eqs. (\ref{pj}) and (\ref{q}), the condition $\partial_{r,r}V_{eff}=0$ leads to
\begin{eqnarray}
  4 r^3 \left(a^2M-r^3+3M r^2-3M^2  r\right)\nonumber\\
  -2 \eta  r \left(3 a^2+5 r^2-6 Mr\right)+5 \eta ^2=0. \label{ccond}
\end{eqnarray}
By solving this condition, we can determine the radius of the spherical orbit where the orbit transitions from stability to instability. Interestingly, calculating the extremal point of the reduced angular momentum $\xi$, defined by $\partial_r \xi=0$ yields the same condition as given in (\ref{ccond}). Additionally, the condition $\partial_r \sigma=0$ also corresponds to (\ref{ccond}), but with an additional extremal condition expressed as
\begin{eqnarray}
  2 r^3-6 Mr^2-5\eta=0,\label{ccondb}
\end{eqnarray}
which is independent of the black hole spin. For $\eta=0$, this results in $r=3M$ corresponding precisely to the radius of the photon sphere of the Schwarzschild black hole.

\begin{figure*}
\subfigure[]{\label{Puspyb}\includegraphics[width=4cm]{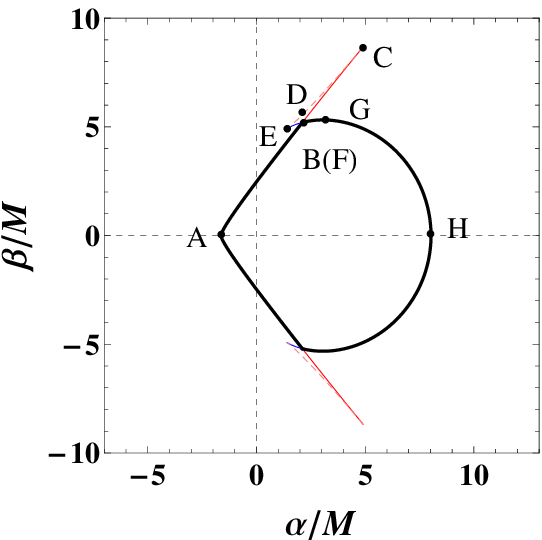}}
\subfigure[]{\label{xii}\includegraphics[width=5cm]{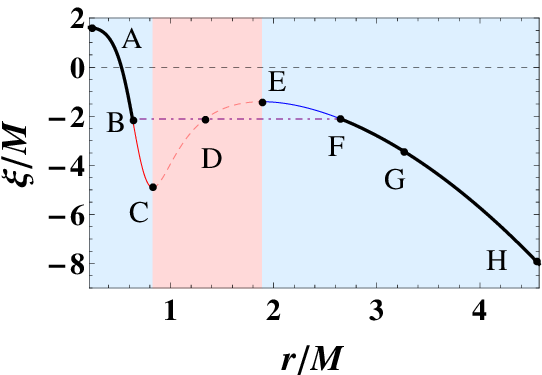}}
\subfigure[]{\label{sigmaa}\includegraphics[width=5cm]{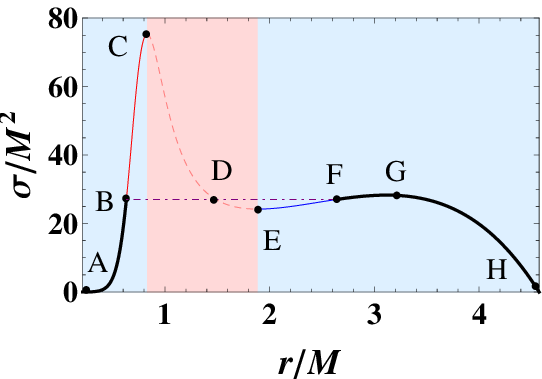}}
\caption{Shadow, reduced angular momentum and Carter constant for the spinning KZ black hole with $a/M=1.6$ and $\eta/M^3=0.45$. The solid and dashed lines correspond to the unstable and stable spherical orbits, respectively. (a) Shadow shape. (b) The reduced angular momentum $\xi$. (c)  The reduced Carter constant $\sigma$. The regions in light blue color and red color are for the unstable and stable spherical orbits. The cusps C and E correspond to the extremal point of $\xi$ and $\sigma$. However, $\sigma$ has an extra extremal point located at point G. Points B and F denote the self-intersection point, where the cuspy shadow connects with the main shadow. Not that these points correspond to different spherical orbits. Points B, D, and F share the same value of $\alpha/M$.}\label{Psigma}
\end{figure*}

To investigate the instability of spherical orbits that underlie the characteristic cuspy shadow of the KZ black hole, we illustrate the shape in Fig. \ref{Puspyb} by taking $a/M=1.6$ and $\eta/M^3=0.45$ as an example. The characteristic points are explicitly indicated. Considering the $\mathcal{Z}_2$ symmetry, we only label them in the upper half plane. The cuspy segments BCEF connect the main shadow at point B or F, corresponding to different spherical orbits. The spherical orbits associated with segments AC and EH are unstable, while segment CE is stable. The reduced angular momentum $\xi$ and the Carter constant $\sigma$ are depicted in Figs. \ref{xii} and \ref{sigmaa}, respectively. The ranges highlighted in light blue (red) indicate $\partial_{r,r}V_{eff}<0(>0)$. From the figures, it is clear that $\xi$ exhibits two extremal points, C and E, while $\sigma$ has three extremal points: C, E, and G. Points C and E mark the transitions in the stability of the spherical orbits, as determined by condition (\ref{ccond}). In contrast, point G arises from condition (\ref{ccondb}), which delineates the extremal point within the shadow boundary shown in Fig. \ref{Puspyb}.

General studies have shown that unstable spherical orbits form the boundary of the shadow, while stable orbits do not. Consequently, the segment CE does not contribute to the realistic shadow boundary. Although the segment BC is unstable, the radius of the corresponding orbits is smaller than that of FG, and thus it also does not provide a realistic boundary. This has been confirmed in Ref. \cite{CunhaHerdeiroRadu}, regardless of whether the observer is located at a finite distance. In contrast to the segment BC, segment EF corresponds to the unstable spherical orbits with larger radii, making it significant for the shadow boundary. It was also noted in Ref. \cite{CunhaHerdeiroRadu} that these effects can result in the appearance of the ``eye lashes". Meanwhile, for a finite distance observer, a ghost shadow edge in the lensing pattern will be evident.

\begin{figure}
\includegraphics[width=7cm]{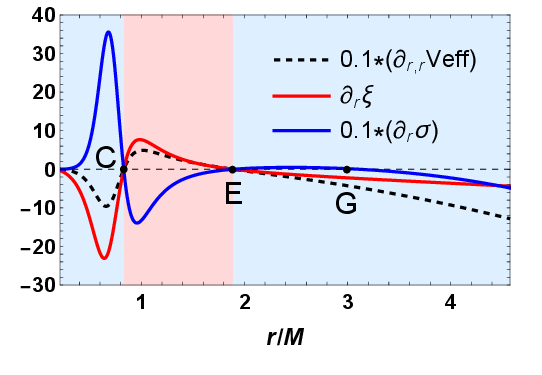}
\caption{The derivatives of $V_{eff}$, $\xi$, and $\sigma$.}\label{PHCDF}
\end{figure}

Let us now focus on the cusps C and E. As demonstrated in Ref. \cite{Weia}, the normal vector of the boundary curve changes by $\pi$ at these points. Here, we provide a reasonable explanation for this phenomenon. We first present three curves for $\partial_r\xi$, and $\partial_r\sigma$ in Fig. \ref{PHCDF}. Notably, segment CE exhibits positive $\partial_{r,r}V_{eff}$, indicating that these correspond to the stable orbits. Specifically, at points C and E, we find $\partial_{r}\xi=0$, $\partial_{r}\sigma=0$ and $\partial_{r,r}V_{eff}=0$. Below or beyond these points, the stability of the spherical orbits transitions. While the slopes of $\xi$ and $\sigma$ vary smoothly, their signs undergo a change. Therefore, in the $\alpha-\beta$ plane, as one approaches $r\to r_{C}^{\pm}$ or $r\to r_{D}^{\pm}$, the slope
\begin{eqnarray}
  \mathcal{F}=\frac{d\beta}{d\alpha}
\end{eqnarray}
keeps a constant. This property results in a reversal of both the tangent and normal vectors of the boundary curve, indicating a sudden change of $\pi$ in their arguments. In contrast to points C and E, although $\partial_{r}\sigma$ changes its sign at point G, $\partial_{r}\xi$ remains smooth and differentiable. Consequently, the boundary curve only alters its direction in the $\beta$ coordinate while maintaining its direction in the $\alpha$ coordinate. As a result, point G is classified as an extremal point rather than a cuspy point. Notably, such slope changes have significant implications for the study of topology and the correspondence between gravity and thermodynamics.

By maintaining the black hole spin at $a/M=1.6$ and increasing $\eta$ such that $\eta\gtrsim1.71M^3$, the reduced angular momentum $\xi$ becomes a monotonically decreasing function of $r$. However, $\sigma$ consistently possesses an extremal point, denoted as point F. Notably, it is important to emphasize that point F will disappear when the black hole spin $a$ becomes sufficiently large.

\section{Topological properties}
\label{Topology}

In this section, we aim to examine the topological properties of the shadow boundary, which are often neglected in shadow studies. Although certain segments of the boundary curve correspond to the stable spherical orbits and do not contribute to a realistic shadow, they remain important for our theoretical analysis.

Reference \cite{CunhaBerti} first considered the topological charge associated with the light ring in the context of compact objects. A study was subsequently generalized to black holes \cite{Cunha}. The result reveals the existence of at least one unstable light ring located outside the outer horizon of the spinning Kerr-like black hole.

For the shadow boundary, the concept of topology was first introduced in Ref. \cite{Weift}, utilizing the Gauss-Bonnet theorem
\begin{eqnarray}
\delta=\frac{1}{2\pi}\left(\oint\frac{dl}{\mathcal{R}}+\sum_i\Delta\theta_i\right),\label{jif}
\end{eqnarray}
where the first term measures the integral over the smooth parts of the boundary, while the second term accounts for the discrete sum over the exterior angles $\Delta\theta_i$ at all non-differentiable cuspy points. Utilizing the celestial coordinates given in Eqs. (\ref{aaa}) and (\ref{bbb}), the local radius of curvature $\mathcal{R}$ is expressed as follows:
\begin{eqnarray}
  \mathcal{R}=\frac{\sqrt{(\alpha'^2+\beta'^2)^3}}{\alpha''\beta'-\alpha'\beta''}. \label{koap}
\end{eqnarray}
Unlike the perimeter and area of the shadow, the local radius of curvature $\mathcal{R}$ varies at different points along the boundary curve, providing detailed local properties for a given shadow \cite{Weift, WeiZou, Omwoyo}. By considering certain characteristic points of the shadow boundary, we can use this information to constrain the black hole parameters based on observed results. Interestingly, as noted in Ref. \cite{WeiZou}, it is possible to estimate the black hole parameters using just a single point, the top point, of the shadow boundary. This approach offers a unique method for testing the characteristics of a black hole without requiring a complete understanding of the black hole shadow. For the D-shape shadow, $\mathcal{R}$ defined here is positive; however, for a cuspy shadow, it could be negative.

In the case of the D-shape shadow, the second term in Eq. (\ref{koap}) vanishes, thus allowing us to evaluate only the first term. Using the curvature $\mathcal{R}$, the topological charge is calculated for the shadow cast by Kerr spacetime with varying spins. The results indicate that for Kerr black holes with different spins, we always have $\delta=1$, whereas $\delta<1$ for Kerr naked singularities, suggesting that the shadow is no longer closed \cite{Weift}.

\begin{figure}
\subfigure[$a/M=0.9$]{\label{Vectoras}\includegraphics[width=4cm]{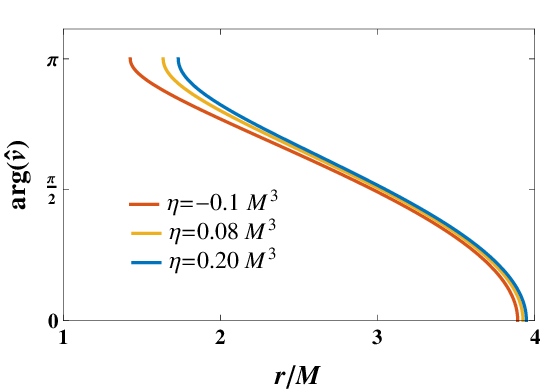}}
\subfigure[$a/M=1.2$]{\label{VectoraL}\includegraphics[width=4.5cm]{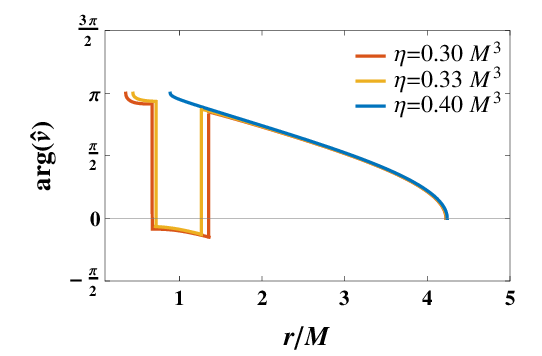}}
\caption{The argument of the tangent vector $v$. (a) $a/M=0.9$. (b) $a/M=1.2$. A sudden change $\pi/2$ of the argument occurs for case (b) with $\eta/M^3$=0.30 and 0.33. }\label{PVectoras}
\end{figure}

On the other hand, the topological charge can be calculated using the tangent vector $v$ of the shadow boundary, defined as
\begin{eqnarray}
  v=(\partial_r \alpha, \partial_r\beta).
\end{eqnarray}
We present the argument of the tangent vector $v$ in Fig. \ref{PVectoras} for $a/M=0.9$ and 1.2. For a small black hole spin of $a/M$=0.9,  as $r$ varies from its minimum to maximum values, $arg(v)$ decreases monotonically from $\pi/2$ to 0. Considering the $\mathcal{Z}_2$ symmetry, the topological charge is \cite{Weia}
\begin{eqnarray}
 \delta=1.
\end{eqnarray}
Alternatively, the charge can also be calculated as
\begin{eqnarray}
 \delta&=&\frac{1}{\pi}\left(\arctan \mathcal{F}(r_{b})-\arctan \mathcal{F}(r_{a})\right)\nonumber\\
 &=&\frac{1}{\pi}\left(\arctan (\infty)-\arctan (-\infty)\right)\nonumber\\
 &=&1.
\end{eqnarray}
Note that if the spin $a$ is sufficiently large, the boundary curve of the shadow will not be smooth at points A and H, resulting in violations of $\arctan \mathcal{F}(r_{a})=-\infty$ and $\arctan \mathcal{F}(r_{a})=\infty$. However, the topological charge remains unchanged.

For a highly spinning black hole with $a/M=1.2$, we observe that for $\eta=0.4 M^3$, the results are similar to those for $a/M=0.9$. However, for $\eta=0.3 M^3$ and $0.33M^3$, the argument $arg(v)$ becomes discontinuous. Each case exhibits two discontinuous points, which are associated with the two cusps of the shadows. A detailed calculation reveals that the changes in $arg(v)$ are exactly equal to $\pi$. This can be readily understood by considering the vector $v$. At the cuspy points, the values of $\partial_r\xi\sim\partial_r\alpha$ and $\partial_r\sigma\sim\partial_r\beta$ remain the same while changing their signs (see Fig. \ref{PHCDF}), leading to
\begin{eqnarray}
 v\to-v.
\end{eqnarray}
As a result, the argument of $v$ experiences a change of $\pi$.

\begin{figure}
\subfigure[D-shape shadow]{\label{Toposhape}\includegraphics[width=4cm]{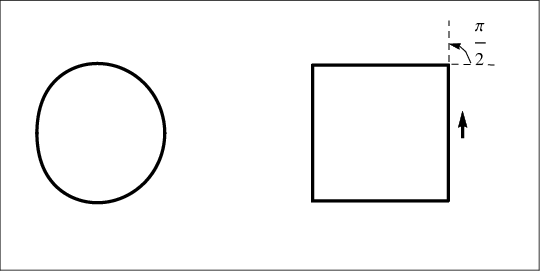}}
\subfigure[Cuspy shadow]{\label{Toposhapeei}\includegraphics[width=4cm]{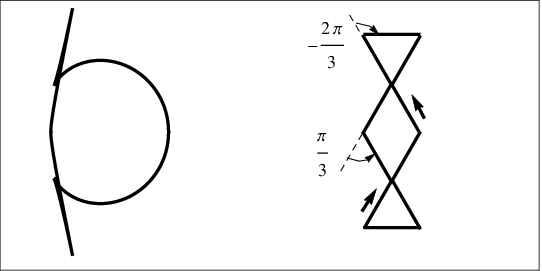}}
\caption{Topological sketch pictures for the D-shape and cuspy shadows. (a) D-shape shadow. (b) Cuspy shadow.}\label{PToposhapeei}
\end{figure}

On the other hand, we observe from Eq. (\ref{jif}) that both the curvature term and the exterior angle term contribute to the topological charge. If one continuously varies the contributions of these terms, the charge remains unchanged. Considering this, we can represent the shadow boundary curves in two characteristic patterns, as shown in Fig. \ref{PToposhapeei}. The D-shape shadow can smoothly and continuously transition to a rectangle, as illustrated in Fig. \ref{Toposhape}. Since each exterior angle takes the value $\pi/2$, we can easily compute
\begin{eqnarray}
 \delta=\frac{1}{2\pi}\left(4\times \frac{\pi}{2}\right)=1,
\end{eqnarray}
which precisely confirms the result shown in Fig. \ref{PVectoras} for the D-shape shadows. For the cuspy shadow, we can represent it as another pattern, specifically an 8-shape, in Fig. \ref{Toposhapeei}. We take the counterclockwise direction as positive, allowing for both positive and negative exterior angles. For this pattern, which consists of six exterior angles, we can express the topological charge as
\begin{eqnarray}
 \delta=\frac{1}{2\pi}\left(2\times \frac{\pi}{3}-4\times\frac{2\pi}{3}\right)=-1,
\end{eqnarray}
confirming the findings of Ref. \cite{Weia} for the cuspy shadow. Clearly, the D-shape shadow and the cuspy shadow possess different values of topological charges, indicating that they belong to distinct topological classifications. This distinctions align perfectly with the results presented in Ref. \cite{Weia}. For other interesting shadows, we can expect the emergence of different topological shapes.

\section{Gravity/thermodynamics correspondence}
\label{analogy}

As demonstrated in Figs. \ref{Efacd5}-\ref{Efccd5}, we observe that the cuspy pattern diminishes as $\eta$ increases, ultimately vanishing when $\eta$ approaches a specific critical value. This behavior closely resembles the swallowtail phenomena observed in the Gibbs free energy $G$ of charged AdS black holes \cite{Kubiznak,Weitwp}.

\begin{figure}
\label{Gibbs}\includegraphics[width=7cm]{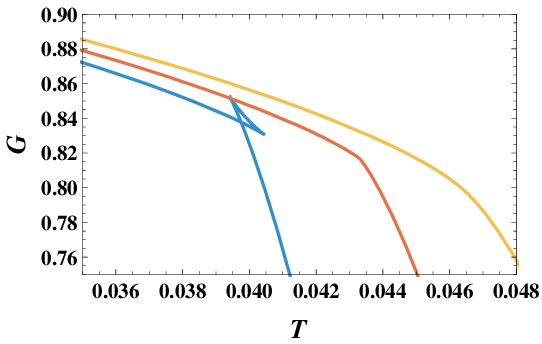}
\caption{Gibbs free energy $G$ for the charged AdS black hole with different pressures. From left to right, the pressure $P=0.8P_c$, $P_c$, and $1.2 P_c$ with $P_c$ the critical pressure. For $P=0.8P_c$, a swallow tail behavior is exhibited, similar to the cuspy shadow. At $P=P_c$, the behavior just disappears. When $P>P_c$, $G$ is a monotonic function of $T$.}\label{PTGibbs}
\end{figure}

For a clear comparison, we present the characteristic behavior of the Gibbs free energy in Fig. \ref{PTGibbs}. This figure illustrates the phase transition between small and large black holes. The differential law for the free energy is given by
\begin{eqnarray}
 dG=-SdT+VdP+\mu dQ,\label{gggg}
\end{eqnarray}
where $T$, $P$, and $Q$ represent the temperature, pressure, and charge of the black hole system, respectively. The corresponding conjugate quantities $S$, $V$, and $\mu$ are the entropy, volume, and chemical potential. The phase transition point is precisely located at the self-intersection point of the swallowtail behavior, denoted by
\begin{eqnarray}
 \Delta G=0.
\end{eqnarray}
In particular, near the critical point determined by
\begin{eqnarray}
 \frac{\partial T}{\partial S}=0,\quad \frac{\partial^2 T}{\partial S^2}=0,
\end{eqnarray}
where the swallowtail behavior disappears, critical phenomena emerge. There are various methods to determine the phase transition point and the critical point, which we will explore later.

Given their striking similarity, it is worthwhile to perform a thermodynamic-like construction for the cuspy shadow, as this may reveal the underlying gravitational properties in the vicinity of the black hole horizon.

From the previous discussion, we know that these two celestial coordinates are functions of $r$, $\eta$, and $a$ after being scaled by the mass. By solving for $r$ from Eq. (\ref{aaa}), we obtain $r=r(\alpha, \eta,  a)$. Substituting this into Eq. (\ref{bbb}), we have $\beta=\beta(r(\alpha, \eta, a), \eta, a)$. Thus, we can derive the following differential expression for $\beta$:
\begin{eqnarray}
 d\beta=\mathcal{F} d\alpha+\mathcal{A} da+\Theta d\eta,\label{bebe}
\end{eqnarray}
where
\begin{eqnarray}
 \mathcal{F}&=&\frac{\partial\beta}{\partial r}\frac{\partial r}{\partial \alpha},\\
 \mathcal{A} &=&\frac{\partial\beta}{\partial r}\frac{\partial r}{\partial a}+\frac{\partial\beta}{\partial a},\\
 \Theta &=&\frac{\partial\beta}{\partial r}\frac{\partial r}{\partial \eta}+\frac{\partial\beta}{\partial \eta}.
\end{eqnarray}
By comparing Eqs. (\ref{gggg}) and (\ref{bebe}), we can establish the correspondence between thermodynamics and gravity. For convenience, we summarize these correspondences in Table \ref{tab1}. In particular, the celestial coordinate $\beta$ acts as the Gibbs free energy, while $\alpha$ serves as the temperature. The slope $\mathcal{F}$ represents the negative entropy of the system. These quantities warrant further exploration.

\begin{table}[h]
\begin{center}
\begin{tabular}{cc|cc}
  \hline\hline
     \multicolumn{2}{c|}{gravity} & \multicolumn{2}{c}{thermodynamics} \\
\hline
orbit radius &$r$&$r_{h}$& horizon radius\\
celestial coordinate &$\beta$&$G$& free energy\\
celestial coordinate &$\alpha$&$T$& temperature\\
slope &$\mathcal{F}$&-$S$& entropy\\
spin &$a$&$P$& pressure\\
deformation parameter  &$\eta$&$Q$& charge\\
conjugate quantity &$\mathcal{A}$&$V$& volume\\
conjugate quantity &$\Theta$&$\mu$& chemical potential\\
self-intersection point & & & phase transition point\\
 \hline\hline
\end{tabular}
\caption{Gravity/thermodynamics correspondence.}\label{tab1}
\end{center}
\end{table}

\section{Self-intersection point}
\label{point}

In this section, we will explore the self-intersection point of the shadow boundary, which marks the location where the cuspy shadow shape connects with the main shadow. From a thermodynamic perspective, this point corresponds to the phase transition point at which two systems coexist. Thus, we refer to it as the geometric phase transition point. There are several methods to determine this point, which we will examine in the following discussion.

\subsection{Cuspy behavior}

The cuspy shapes correspond to the swallowtail behaviors of the thermodynamic free energy. Thus, the first method to obtain the self-intersection point is specified by the condition
\begin{eqnarray}
 \Delta \beta=0,\label{ddbb}
\end{eqnarray}
at two radii $r_1$ and $r_2$ of the spherical orbits, where we generally assume $r_2>r_1$. The condition (\ref{ddbb}) simplifies to
\begin{eqnarray}
 \alpha(r_1, \eta, a)=\alpha(r_2, \eta, a),\quad \beta(r_1, \eta, a)=\beta(r_2, \eta, a).
\end{eqnarray}
For given values of $a$ and $\eta$, we can determine $r_1$ and $r_2$ by solving these two equations.

\begin{figure}
\subfigure[]{\label{ptea}\includegraphics[width=4cm]{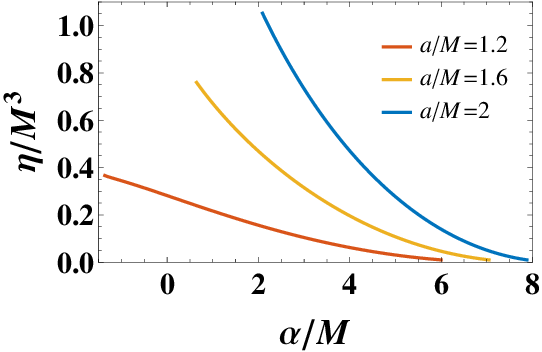}}
\subfigure[]{\label{pterr}\includegraphics[width=4cm]{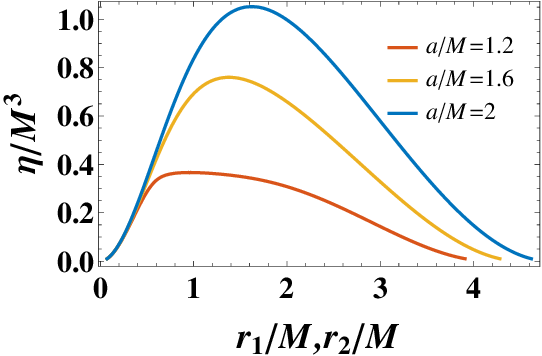}}
\caption{The self-intersection curves. (a) $\eta/M^3-a/M$ plane. (b)  $\eta/M^3-r/M$ plane. $a/M$=1.2, 1.6, and 2 from bottom to top. The top point of each curve corresponds to the critical point. The corresponding critical values of $\eta_c$=0.3664, 0.7597, 1.0520.}\label{PTpterr}
\end{figure}

In Fig. \ref{PTpterr}, we present the numerical results for the self-intersection point curve, varying $a/M$ among 1.2, 1.6, and 2.0. It is evident that for each given spin, $\eta$ decreases with $\alpha$, as illustrated in Fig. \ref{ptea}. Unlike the temperature, $\alpha$ can take negative values. In Fig. \ref{pterr}, we observe that the critical value of $\eta$ increases with the black hole spin. The behavior of the self-intersection point curve closely resembles the coexistence curve of the black hole phase transition. The regions below each curve represent the coexistence region where small and large black holes coexist; however, in this context, it pertains to the stable spherical orbits corresponding to the geometric phase transition.

\subsection{Gravitational equal-area law}

As is well known, the phase transition point can also be equivalently determined using Maxwell's equal-area law. Similarly, the self-intersection point of the cuspy shadow can be established by constructing an equal-area law. In Ref. \cite{Weia}, a gravitational equal-area law was proposed, and we aim to clarify this further.

\begin{figure}
\subfigure[]{\label{Eqarlaw}\includegraphics[width=4cm]{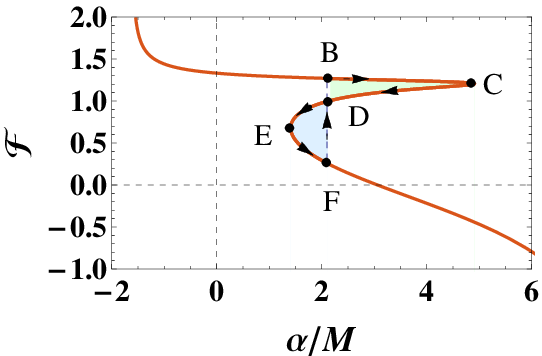}}
\subfigure[]{\label{Eqalab}\includegraphics[width=4cm]{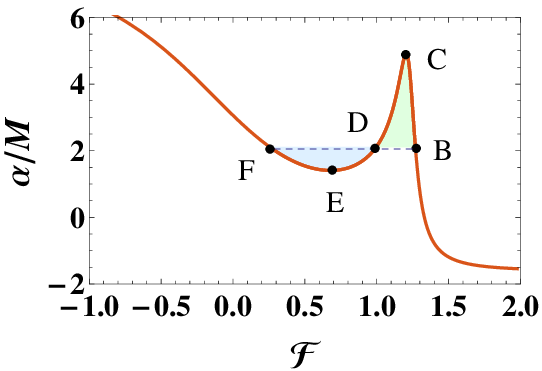}}
\caption{Gravitational equal-area law. (a) $\mathcal{F}$-$\alpha/M$ plane. (b) $\alpha/M$-$\mathcal{F}$ plane. These two areas in light green and blue color are equal. These points B, C,and D et al correspond to these shown in Fig. \ref{Psigma}.} \label{PEqalab}
\end{figure}

Considering the spherical orbits with different radii $r_1$ and $r_2$ corresponding to the same self-intersection point, we have $\beta_1=\beta_2$. Thus, the differential formula for $\beta$ takes the following form
\begin{eqnarray}
 \mathcal{F}d\alpha+\mathcal{A} da+\Theta d\eta=0.
\end{eqnarray}
Integrating it, we find
\begin{eqnarray}
 \int_{\alpha_1}^{\alpha_2}\mathcal{F}d\alpha+\int_{a_1}^{a_2}\mathcal{A} da+\int_{\eta_1}^{\eta_2}\Theta d\eta=0.
\end{eqnarray}
By further fixing the deformation parameter $\eta$ and spin $a$, the above equation simplifies to
\begin{eqnarray}
 \int_{\alpha_1}^{\alpha_2}\mathcal{F}d\alpha=0.
\end{eqnarray}
At the self-intersection point, these two spherical orbits share the same $\alpha$. Consequently, we establish the gravitational equal-area law
\begin{eqnarray}
 \oint\mathcal{F}d\alpha=0.
\end{eqnarray}
Following the closed path BCDEFDB depicted in Fig. \ref{Eqarlaw}, this formula may not be convenient for specific calculations. To address this, we can change the integration variable to $\mathcal{F}$
\begin{eqnarray}
 \int_{\mathcal{F}_1}^{\mathcal{F}_2}\alpha d\mathcal{F}=\alpha_*(\mathcal{F}_2-\mathcal{F}_1),\label{eqarea}
\end{eqnarray}
where $\mathcal{F}_2$ and $\mathcal{F}_1$ correspond to the same $\alpha_*$ at the self-intersection point. Here, $\alpha$ is a function of $\mathcal{F}$, $\eta$, and $a$. The left-hand side measures the area under the curve $\alpha$ bounded by $\mathcal{F}_1$ and $\mathcal{F}_2$, while the right-hand side denotes the area of a rectangle with width $\alpha_*$ and length $(\mathcal{F}_2-\mathcal{F}_1)$. If Eq. (\ref{eqarea}) holds true, the two areas, BCDB (shaded in light green) and DEFD (shaded in light blue), enclosed by the curve $\alpha$ and the horizontal line $\alpha=\alpha_*$ are equal. This relationship corresponds to the equal-area law and is also valid for the cuspy shadow. By performing specific calculations, we can obtain these results consistent with those shown in Figs. \ref{ptea} and \ref{pterr}.

\subsection{$\tilde{\beta}$-landscape}

Here, we introduce a third method for determining the self-intersection point.

Through our gravity/thermodynamics correspondence, we can define the gravitational enthalpy by performing a Legendre transformation
\begin{eqnarray}
 \mathcal{H}=\beta-\alpha \mathcal{F}.
\end{eqnarray}
Building upon this, we can define a generalized celestial coordinate
\begin{eqnarray}
 \tilde{\beta}=\mathcal{H}+\tilde{\alpha}\mathcal{F}=\beta-(\alpha-\tilde{\alpha}) \mathcal{F},
\end{eqnarray}
where $\tilde{\alpha}$ is a constant. It is important to note that only when $\alpha=\tilde{\alpha}$ does $\tilde{\beta}=\beta$. These celestial coordinates reside on the shadow boundary, representing a viable physical solution.

Let us differentiate $\tilde{\beta}$ with respect to $r$
\begin{eqnarray}
 \frac{\partial{\tilde{\beta}}}{{\partial r}}&=&
 \frac{\partial{\beta}}{\partial r}
 -\frac{\partial \alpha}{\partial r}\mathcal{F}
 -(\alpha-\tilde{\alpha})\frac{\partial\mathcal{F}}{\partial r}\nonumber\\
 &=&(\tilde{\alpha}-\alpha)\frac{\partial\mathcal{F}}{\partial r}.
\end{eqnarray}
In general, we have $\frac{\partial\mathcal{F}}{\partial r}\neq0$, leading to the condition $\frac{\partial{\tilde{\beta}}}{{\partial r}}=0$ when $\alpha=\tilde{\alpha}$. Consequently, the physically realistic solution occurs at the extremal point of  $\tilde{\beta}$. At this point, we further find
\begin{eqnarray}
 \frac{\partial^2{\tilde{\beta}}}{{\partial r^2}}
 =-\frac{\partial\alpha}{\partial r}\frac{\partial\mathcal{F}}{\partial r}.
\end{eqnarray}
In general, since $\frac{\partial\mathcal{F}}{\partial r}<0$, we have
\begin{eqnarray}
 \frac{\partial^2{\tilde{\beta}}}{{\partial r^2}}
 \propto \frac{\partial\alpha}{\partial r}
 \sim -\frac{\partial\xi}{\partial r}.
\end{eqnarray}
From this analysis, we conclude that the unstable and stable spherical orbits correspond to negative and positive values of  $(\partial_r\xi)$, respectively. Therefore, the maximum and minimum points of $\tilde{\beta}$ correspond to stable and unstable orbits.

\begin{figure*}
\subfigure[$\tilde{\alpha}/M$=1.8]{\label{Landscapeab}\includegraphics[width=5cm]{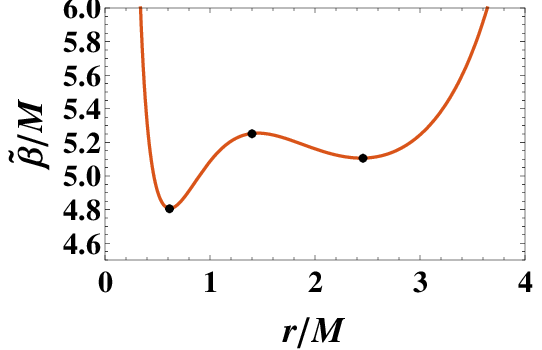}}
\subfigure[$\tilde{\alpha}/M$=2.11]{\label{Landscapeabb}\includegraphics[width=5cm]{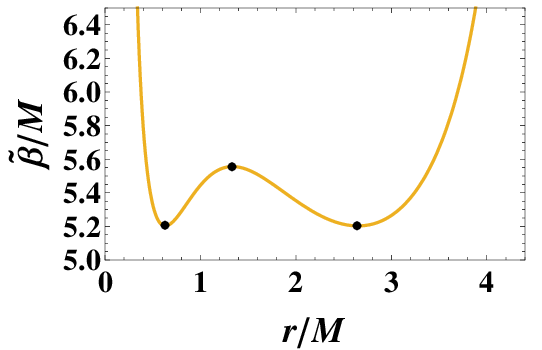}}
\subfigure[$\tilde{\alpha}/M$=3]{\label{Landscapeacc}\includegraphics[width=5cm]{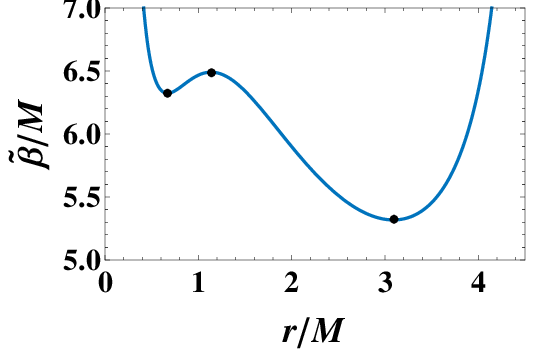}}
\caption{$\tilde{\beta}$-landscapes for the spinning KZ black hole with $a/M=1.6$ and $\eta/M^3=0.45$. (a) $\tilde{\alpha}/M$=1.8. (b) $\tilde{\alpha}/M$=2.11. (c) $\tilde{\alpha}/M$=3. The extremal points marked in black color denote the physical spherical orbits. The minimum and maximum points correspond to unstable and stable ones.}\label{PTLandscapeacc}
\end{figure*}

Taking $a/M=1.6$ and $\eta/M^3=0.45$, we present the generalized celestial coordinate $\tilde{\beta}$ in Fig. \ref{PTLandscapeacc} for $\tilde{\alpha}=$1.8, 2.11, and 3, respectively. In each case, there are three extremal points: one maximum and two minimum points. For small values of $\tilde{\alpha}$, the left well is deeper than the right one; however, for large values of $\tilde{\alpha}$, this situation is reversed. Notably, there exists a special case where the two wells have the same depth, which corresponds exactly to the self-intersection point.

In the context of the $\tilde{\beta}$-landscape study, the self-intersection points $r_1$ and $r_2$ can be determined by the following conditions
\begin{eqnarray}
 \tilde{\beta}(r_1)=\tilde{\beta}(r_2),\quad
 \tilde{\beta}'(r_1)=\tilde{\beta}'(r_2)=0.
\end{eqnarray}
By employing this approach, we can also derive the self-intersection points for the black holes with different spins. The results align precisely with those presented in Fig. \ref{PTpterr}.

So far, we have introduced three methods to determine the self-intersection point, or geometric phase transition point. Notably, these methods are found to be equivalent.

\section{Critical phenomena}
\label{phenomena}

In the previous section, we examined the self-intersection point of the shadow boundary, which corresponds to the phase transition of a thermodynamic system. Here, we aim to explore the critical phenomena near the critical point, where the cuspy pattern either appears or disappears.

\subsection{Critical point}

As a first step, we aim to determine the critical point. As shown above, the two cusps occur at the extremal points of $\alpha$ or $\xi$ for a given black hole spin. As $\eta$ increases, these extremal points approach and ultimately merge at a critical point, where the radii of the two spherical orbits coincide. Therefore, we can utilize the following conditions
\begin{eqnarray}
 \frac{\partial \xi}{\partial r}=0,\quad \frac{\partial^2 \xi}{\partial r^2}=0,
 \label{cpcpp}
\end{eqnarray}
to obtain the critical point of this geometric phase transition. Additionally, we can apply any of the three methods mentioned in Sec. \ref{point} to arrive at the same result. By solving Eq. (\ref{cpcpp}), we derive the analytical results
\begin{eqnarray}
 a_{c}/M&=&\frac{1}{2}\sqrt{8\sqrt{\frac{10\eta}{M^3}}-50\frac{\eta}{M^3}+25\sqrt{\frac{5\eta^{3}}{2M^9}}},\\
 r_{c}/M&=&\sqrt{\frac{5\eta}{2M^3}}.
\end{eqnarray}
In the $\eta/M^3$-$a/M$ parameter space, the value starts at zero and gradually increases with the black hole spin for $a/M<1$. When $a/M>1$, it exhibits a rapid increase, extending to large values of $\eta/M^3$ and $a/M$. However, requiring the spherical orbit to lie outside the black hole horizon yields a minimal critical spin value of $a_{c}^{min}/M=1.13$, which exceeds the Kerr limit. This also implies that the cuspy shadow is absent for the Kerr black holes.

\subsection{Critical exponent}

The critical exponent is another important parameter for testing critical phenomena. In Ref. \cite{Weia}, a numerical study for $a/M$=2 indicates that $\Delta r=r_2-r_1$ and $\Delta \mathcal{F}=\mathcal{F}_2-\mathcal{F}_1$ yield an approximate critical exponent value of $1/2$. This raises the question of whether an analytical study could produce the similar results.

To address this question, we note that the behavior of the cuspy shadow closely resembles the phase transition between small and large black holes. In this context, it is generally observed that $r_2> r_1$, and at the critical point, $r_2=r_1$. Following Ref. \cite{Weicr}, we can define the ratio
\begin{eqnarray}
 \epsilon=1-\frac{r_1}{r_2}, \quad \epsilon\in[0, 1).
\end{eqnarray}
If we can analytically expand these quantities near $\epsilon=0$, we will obtain the exact exponent. First, let us determine the radii $r_1$ and $r_2$ of the spherical orbits corresponding to the self-intersection point of the shadow boundary. As discussed in Sec. \ref{point}, there are three methods available; however, obtaining $r_1$ and $r_2$ as analytical functions of $a$ and $\eta$ (or $\alpha$) poses unique challenges.

Interestingly, if we substitute $r_1$ with $\epsilon$, such that $r_1=(1-\epsilon)r_2$, these equations can be solved analytically. However, the resulting expressions are quite cumbersome, and we omit them here. Near $\epsilon=0$, we perform the following expansions
\begin{eqnarray}
 \frac{\eta}{M^3}&=&0.7597-0.1778 \epsilon^2+\mathcal{O}(\epsilon^3),\\
 \frac{\Delta r}{M}&=&1.3782\epsilon+0.6891\epsilon^2+\mathcal{O}(\epsilon^3),
\end{eqnarray}
for $a/M=1.6$. At the critical point, we clearly find that $\Delta r/M$=0 and $\eta/M^3=0.7597$, which corresponds exactly to the critical value of $\eta$. Utilizing the expansions mentioned above, we can express
\begin{eqnarray}
 \frac{\Delta r}{M}=-3.2683\left(0.7597-\frac{\eta}{M^3}\right)^{\frac{1}{2}}+\mathcal{O}\left(0.7597-\frac{\eta}{M^3}\right)^{\frac{3}{2}}.
\end{eqnarray}
This confirms the following critical exponent
\begin{eqnarray}
 \gamma=\frac{1}{2},
\end{eqnarray}
which is consistent with the numerical result from Ref. \cite{Weia}. Although we have adopted $a/M=1.6$, the exponent remains unchanged for arbitrary values of spin $a$ when considering the analytical expressions for $\eta$ and $r_2$. It is noteworthy that we also have
\begin{eqnarray}
 \epsilon\sim \left(0.7597-\frac{\eta}{M^3}\right)^{\frac{1}{2}}.
\end{eqnarray}
As a result, both the difference $\Delta r$ and the ratio $\epsilon$ of $r_1$ and $r_2$ exhibit the same critical exponent. Notably, the ratio $\epsilon$ defined here vanishes at the critical point, allowing $\epsilon$ to serve as an order parameter characterizing this geometric phase transition, a feature not previously observed.

In summary, we have obtained the following analytical result
\begin{eqnarray}
 \Delta r=a_1\left(\frac{\eta_c-\eta}{M^3}\right)^{\frac{1}{2}}+a_2\left(\frac{\eta_c-\eta}{M^3}\right)^{\frac{3}{2}}+\mathcal{O}\left(\frac{\eta_c-\eta}{M^3}\right)^{2}.
\end{eqnarray}
For other values of black hole spin, we present the values of $a_1$, $a_2$ and $\eta_c$ in Table \ref{tab2}. Notably, we observe that $a_1$ is negative, while $a_2$ is positive.

\begin{table}[h]
\begin{center}
\begin{tabular}{cccc}
  \hline\hline
    $a$ &$a_1$&$a_2$& $\eta_c/M^3$\\\hline
1.2 &-6.4910&31.5050& 0.3664\\
1.6 &-3.2683&1.0729& 0.7597\\
2.0 &-2.9824&0.5988& 1.0520\\
3.0 &-2.7676&0.2846& 1.7108\\
5.0 &-2.6443&0.1063& 3.0131\\
 \hline\hline
\end{tabular}
\caption{Values of the expansion coefficients $a_1$, $a_2$, and $\eta_c$.}\label{tab2}
\end{center}
\end{table}

Using this method, other critical exponents analogous to those of thermodynamic phase transitions can be obtained analytically. The results confirm that all the exponents of the cuspy shadow align with those predicted by mean field theory, as detailed in Table \ref{tab1} through our gravity/thermodynamics correspondence.

\section{Discussions and conclusions}
\label{Conclusion}

In this paper, we have examined the topology and the gravity/thermodynamics correspondence through the cuspy shadows cast by spinning KZ black holes. We proposed that all the information underlying the shadow can be revealed by the parametric equation of the shadow boundary.

It is well known that the shadow boundary is formed by unstable spherical orbits, which correspond to the D-shape shadow of the spinning Kerr black holes. In non-Kerr cases, the stable spherical orbits may also exist, leading to the emergence of another characteristic shape: the cuspy shadow. By investigating this feature, we aimed to enhance our understanding of the physics in the gravitational regime beyond general relativity. Here, we used the spinning KZ black hole as an example to thoroughly explore the cuspy shadow and to further develop the gravity/thermodynamics correspondence.

By examining the cuspy shadow in the celestial plane, we investigated the reduced angular momentum $\xi$ and the Carter constant $\sigma$. The results indicate that these two cusps correspond precisely to the extremal points of $\xi$ and $\sigma$. The stable spherical orbit branch is bounded by these two extremal points. Unlike $\xi$, $\sigma$ exhibits an additional extremal point, which results in an extremal point rather than a cusp in the shadow boundary in the celestial plane. Thus, our study establishes a clear relationship between the cusps, the extremal points of the shadow, and the spherical orbits.

Building on this result, we followed Ref. \cite{Weia} to explore the topology of the shadow boundary. For the D-shape shadow, the topological charge is consistently $\delta=1$, as expected. In contrast, the topological charge for the cuspy shadow is -1, attributed to each cuspy feature representing one genus. To clearly illustrate the difference in topology, we depict the D-shape and cuspy shadows as rectangular and 8-shaped topologies in Fig. \ref{PToposhapeei}, highlighting only the exterior angles, which may be either positive or negative. More intriguingly, upon examining the argument of the tangent vector of the cuspy shadow boundary, we observed two sudden changes of $\pi$, each corresponding to one of the two cusps of the shadow. This phenomenon arises from the direction reversal of the tangent vector, characterized by $\partial_r\xi=\partial_r\sigma=\partial_{r,r}V_{eff}=0$, which delineates the boundary between stable and unstable spherical orbits.

By comparing the cuspy shadow with the Gibbs free energy of a thermodynamic system, we established a correspondence between gravity and thermodynamics. The celestial coordinates $\alpha$ and $\beta$ correspond to the temperature and the free energy, respectively. Consequently, the self-intersection point of the cuspy shape and the main shadow acts as the phase transition points between two different phases of the system. Drawing an analogy with thermodynamics, we proposed three methods, cuspy behavior, the gravitational equal-area law, and the $\tilde{\beta}$-landscape, to determine the self-intersection point of this geometric phase transition. Notably, these three methods are equivalent. In particular, the  $\tilde{\beta}$-landscape is a completely new concept that warrants further investigation.

Lastly, the critical exponent is analytically calculated by defining the ratio of the radii of the large and small unstable spherical orbits. The numerical result of 1/2 from Ref. \cite{Weia} is confirmed analytically by our results. This result also suggests that this ratio, akin to the difference in the orbital radii, can serve as an order parameter characterizing the cuspy shadows.

In conclusion, we have proposed a gravity/thermodynamics correspondence, as detailed in Table \ref{tab1}, which arises from our study of cuspy black hole shadows. While this correspondence is clearly established, a deeper understanding of its nature still warrants further investigation. Such a correspondence will offer valuable insights into the strong gravitational effects near the black hole horizon from a thermodynamic perspective.

\section*{Acknowledgements}
We would like to thank Prof. Songbai Chen for the useful discussions. This work was supported by the National Natural Science Foundation of China (Grants No. 12475055, No. 12475056, No. 12247101).


\begin{thebibliography}{99}

\bibitem{Synge}
 J. L. Synge,
 {\em The escape of photons from gravitationally intense stars},
 MNRAS,  \textbf{131}, 463 (1966).

\bibitem{Luminet}
 J. P. Luminet,
 {\em Image of a spherical black hole with thin accretion disk},
 A. A, \textbf{75}, 228 (1979).

\bibitem{Bardeen}
 J. M. Bardeen,
 {\em Timelike and null geodesics in the Kerr metric}, Les Astres Occlus, (1973).

\bibitem{Chandrasekhar}
 S. Chandrasekhar,
 {\em The Mathematical Theory of Black Holes},
  Oxford University Press, New York, (1992).

\bibitem{Hioki}
 K. Hioki and K. I. Maeda,
 {\em Measurement of the Kerr Spin Parameter by Observation of a Compact Object's Shadow},
  Phys. Rev. D \textbf{80}, 024042 (2009), [arXiv:0904.3575 [astro-ph.HE]].

\bibitem{Amarilla}
 L. Amarilla, E. F. Eiroa, and G. Giribet,
 {\em Null geodesics and shadow of a rotating black hole in extended Chern-Simons modified gravity},
  Phys. Rev. D \textbf{81}, 124045 (2010), [arXiv:1005.0607 [gr-qc]].

\bibitem{Nedkova}
 P. G. Nedkova, V. K. Tinchev, and S. S. Yazadjiev,
 {\em Shadow of a rotating traversable wormhole},
  Phys. Rev. D \textbf{88}, 124019 (2013), [arXiv:1307.7647 [gr-qc]].

\bibitem{Atamurotov}
 F. Atamurotov, A. Abdujabbarov, and B. Ahmedov,
 {\em Shadow of rotating non-Kerr black hole},
  Phys. Rev. D \textbf{88}, 064004 (2013).


\bibitem{Wei}
 S.-W. Wei and Y.-X. Liu,
 {\em Observing the shadow of Einstein-Maxwell-Dilaton-Axion black hole},
  JCAP \textbf{11}, 063 (2013), [arXiv:1311.4251 [gr-qc]].

\bibitem{Cunhaz}
 P. V. P. Cunha, C. A. R. Herdeiro, and E. Radu,
 {\em Shadows of Kerr black holes with scalar hair},
  Phys. Rev. Lett. \textbf{115}, 211102 (2015), [arXiv:1509.00021 [gr-qc]].

\bibitem{Ghosh}
 M. Amir and S. G. Ghosh,
 {\em Shapes of rotating nonsingular black hole shadows},
  Phys. Rev. D \textbf{94}, 024054 (2016), [arXiv:1603.06382 [gr-qc]].

\bibitem{Chen}
 M. Wang, S. Chen, and J. Jing,
 {\em Shadows of Bonnor black dihole by chaotic lensing},
  Phys. Rev. D \textbf{97}, 064029 (2018), [arXiv:1710.07172 [gr-qc]].

\bibitem{Guo}
 P.-C. Li, M. Guo, and B. Chen,
 {\em Shadow of a Spinning Black Hole in an Expanding Universe},
  Phys. Rev. D \textbf{101}, 084041 (2020), [arXiv:2001.04231 [gr-qc]].

\bibitem{Tsukamoto}
 N. Tsukamoto,
 {\em Black hole shadow in an asymptotically-flat, stationary, and axisymmetric spacetime: The Kerr-Newman and rotating regular black holes},
  Phys. Rev. D \textbf{97}, 064021 (2018), [arXiv:1708.07427 [gr-qc]].

\bibitem{Battista}
 E. Battista, S. Capozziello, and C.-Y. Chen,
 {\em Shadow signatures and energy accumulation in Lorentzian-Euclidean black holes},
  [arXiv:2601.10806 [gr-qc]].
  
\bibitem{GaoHu}
X.-J. Gao, T.-T. Sui, X.-X. Zeng, Y.-S. An, and Y.-P.Hu,
{\em Investigating shadow images and rings of the charged Horndeski black hole illuminated by various thin accretions},
 Eur. Phys. J. C \textbf{83}, 1052 (2023),
[arXiv:2311.11780 [gr-qc]].  

\bibitem{CunhaHerdeiroa}
 P. V. P. Cunha and C. A. R. Herdeiro,
 {\em Shadows and strong gravitational lensing: a brief review},
  Gen. Rel. Grav. \textbf{50}, 42 (2018), [arXiv:1801.00860 [gr-qc]].

\bibitem{Perlick}
 V. Perlick and O. Yu. Tsupko,
 {\em Calculating black hole shadows: Review of analytical studies},
  Phys. Rept. \textbf{947}, 1 (2022), [arXiv:2105.07101 [gr-qc]].

\bibitem{JingQian}
 S. Chen, J. Jing, W.-L. Qian, and B. Wang ,
 {\em Black hole images: A review},
  Sci. China Phys. Mech. Astron. \textbf{66}, 260401 (2023), [arXiv:2301.00113 [astro-ph.HE]].

\bibitem{EHT}
 The Event Horizon Telescope Collaboration,
 {\em First M87 Event Horizon Telescope Results. I. The Shadow of the Supermassive Black Hole},
  Astrophys. J. Lett. \textbf{875}, L1 (2019), [arXiv:1906.11238 [astro-ph.GA]].

\bibitem{EHTb}
 The Event Horizon Telescope Collaboration,
 {\em First Sagittarius A* Event Horizon Telescope Results. I. The Shadow of the Supermassive Black Hole in the Center of the Milky Way},
  Astrophys. J. Lett. \textbf{930 }, L12 (2022), [arXiv:2311.08680 [astro-ph.HE]].

\bibitem{EHTc}
 The Event Horizon Telescope Collaboration,
 {\em Constraints on black-hole charges with the 2017 EHT observations of M87*},
      Phys. Rev. D \textbf{103}, 104047 (2021), [arXiv:2105.09343 [gr-qc]].

\bibitem{EHTd}
 The Event Horizon Telescope Collaboration,
 {\em Horizon-scale variability of M87* from 2017-2021 EHT observations},
      Astron. Astrophys. \textbf{704}, A91 (2025), [arXiv:2509.24593 [astro-ph.HE]].

\bibitem{Gralla}
 S. E. Gralla, D. E. Holz, and R. M. Wald,
 {\em Black Hole Shadows, Photon Rings, and Lensing Rings},
      Phys. Rev. D \textbf{100}, 024018 (2019), [arXiv:1906.00873 [astro-ph.HE]].

\bibitem{Freese}
 C. Bambi, K. Freese, S. Vagnozzi, and L. Visinelli,
 {\em Testing the rotational nature of the supermassive object M87* from the circularity and size of its first image},
      Phys. Rev. D \textbf{100}, 044057 (2019), [arXiv:1904.12983 [gr-qc]].

\bibitem{Zinhailo}
 R. A. Konoplya and O. S. Stashko,
 {\em Probing the effective quantum gravity via quasinormal modes and shadows of black holes},
         Phys. Rev. D \textbf{111}, 104055 (2025), [arXiv:2408.02578 [gr-qc]].

\bibitem{SenGupta}
 I. Banerjee, S. Chakraborty, and S. SenGupta,
 {\em Silhouette of M87*: A New Window to Peek into the World of Hidden Dimensions},
     Phys. Rev. D \textbf{101}, 041301 (2020), [arXiv:1909.09385 [gr-qc]].

\bibitem{Wangb}
     F. Ahmed, H. Ali, Q. Wu, T. Zhu, and S. G. Ghosh,
 {\em Shadows of rotating non-commutative Kiselev black holes: constraints from EHT observations of M87* and Sgr A*},
         Eur. Phys. J. C \textbf{85}, 795 (2025).

\bibitem{Zhangb}
 X.-X. Zeng, C.-Y. Yang, M. I. Aslam, R. Saleem, and S. Aslam,
 {\em Kerr-like Black Hole Surrounded by Cold Dark Matter Halo: The Shadow Images and EHT Constraints},
 JCAP \textbf{08}, 066 (2025), [arXiv:2505.07063 [gr-qc]].

\bibitem{Junior}
 P. V. P. Cunha, C. A. R. Herdeiro, and J. P. A. Novo,
 {\em Light rings on stationary axisymmetric spacetimes: blind to the topology and able to coexist},
        Phys. Rev. D \textbf{109}, 064050 (2024), [arXiv:2401.05495 [gr-qc]].

\bibitem{Hou}
 Y. Hou, Z. Zhang, H. Yan, M. Guo, and B. Chen,
 {\em Image of a Kerr-Melvin black hole with a thin accretion disk},
    Phys. Rev. D \textbf{106}, 064058 (2022), [arXiv:2206.13744 [gr-qc]].

\bibitem{Kuangb}
 X.-M. Kuang, Y. Meng, E. Papantonopoulos, and X.-J. Wang,
 {\em Using the shadow of a black hole to examine the energy exchange between axion matter and a rotating black hole},
      Phys. Rev. D \textbf{110}, L061503 (2024), [arXiv:2406.11932 [gr-qc]].

\bibitem{Yangb}
 J. Yang, C. Zhang, and Y. Ma,
 {\em Shadow and stability of quantum-corrected black holes},
    Eur. Phys. J. C \textbf{83}, 619 (2023), [arXiv:2211.04263 [gr-qc]].

\bibitem{CaoLi}
 L.-M. Cao, L.-Y. Li, and X.-Y. Liu,
 {\em Image of Quantum Improved Regular Kerr Black Hole and Parameter Constraints from EHT Observations},
    Eur. Phys. J. C \textbf{85}, 944 (2025), [arXiv:2410.15745 [gr-qc]].

\bibitem{CunhaHerdeiroRadu}
 P. V. P. Cunha, C. A. R. Herdeiro, and E. Radu,
 {\em Fundamental photon orbits: black hole shadows and spacetime instabilities},
  Phys. Rev. D \textbf{96}, 024039 (2017),
   [arXiv:1705.05461 [gr-qc]].

\bibitem{CunhaBerti}
 P. V. P. Cunha, E. Berti, and C. A. R. Herdeiro,
 {\em Light ring stability in ultra-compact objects},
  Phys. Rev. Lett. \textbf{119}, 251102 (2017),
   [arXiv:1708.04211 [gr-qc]].

\bibitem{Cunha}
 P. V. P. Cunha and C. A. R. Herdeiro,
 {\em Stationary black holes and light rings},
  Phys. Rev. Lett. \textbf{124}, 181101 (2020),
   [arXiv:2003.06445 [gr-qc]].

\bibitem{Weisw}
 S.-W. Wei,
 {\em Topological Charge and Black Hole Photon Spheres},
  Phys. Rev. D \textbf{102}, 064039 (2020),
   [arXiv:2006.02112 [gr-qc]].

\bibitem{WangChen}
 M. Wang, S. Chen, and J. Jing,
 {\em Shadow casted by a Konoplya-Zhidenko rotating non-Kerr black hole},
  JCAP \textbf{10}, 051 (2017),
   [arXiv:1707.09451 [gr-qc]].
   
\bibitem{GyulchevYazadjiev}
G. Gyulchev, P. Nedkova, V. Tinchev, and S. Yazadjiev,
{\em On the shadow of rotating traversable wormholes},
Eur. Phys. J. C \textbf{78}, 544 (2018),
[arXiv:1805.11591 [gr-qc]].   

\bibitem{Qian}
 W.-L. Qian, S. Chen, C.-G. Shao, B. Wang, and R.-H. Yue,
 {\em Cuspy and fractured black hole shadows in a toy model with axisymmetry},
 Eur. Phys. J. C \textbf{82}, 91 (2022), [arXiv:2102.03820 [gr-qc]].

\bibitem{Ohta}
 C.-Y. Chen, C.-M. Chen, and N. Ohta,
 {\em Shadow of black holes with consistent thermodynamics},
  [arXiv:2510.00708 [gr-qc]].

\bibitem{ChengZhao}
 P. Cheng, R.-F. Xu, and P. Zhao,
 {\em On the Cuspy Structure of Rotating Wormhole Shadows},
  [arXiv:2602.14182 [gr-qc]].
  
\bibitem{ChengYang}
P. Cheng and S.-J. Yang,
{\em On the Universal Cuspy Behavior in Black Hole Shadows},
[arXiv:2603.19576 [gr-qc]].  

\bibitem{Weia}
 S.-W. Wei, C.-H. Wang, Y.-P. Zhang, Y.-X. Liu, and R. B. Mann,
 {\em Gravitational equal-area law and critical phenomena of cuspy black hole shadow},
   [arXiv:2601.15612 [gr-qc]].

\bibitem{KonoplyaZhidenko}
 R. Konoplya and A. Zhidenko,
 {\em Detection of gravitational waves from black holes: Is there a window for alternative theories?},
  Phys. Lett. B \textbf{756}, 350 (2016),
   [arXiv:1602.04738 [gr-qc]].

\bibitem{Weift}
 S.-W. Wei, Y.-X. Liu, and R. B. Mann,
 {\em Intrinsic curvature and topology of shadows in Kerr spacetime},
  Phys. Rev. D \textbf{99}, 041303 (2019),
   [arXiv:1811.00047 [gr-qc]].

\bibitem{WeiZou}
 S.-W. Wei, Y.-C. Zou, Y.-X. Liu, and R. B. Mann,
 {\em Curvature radius and Kerr black hole shadow},
  JCAP \textbf{1908}, 030 (2019),
   [arXiv:1904.07710 [gr-qc].

\bibitem{Omwoyo}
 E. Omwoyo, H. Belich, J. C. Fabris, and H. Velten,
 {\em Remarks on the black hole shadows in Kerr-de Sitter space times},
  Eur. Phys. J. C \textbf{82}, 395 (2022),
   [arXiv:2112.14124 [gr-qc]].

\bibitem{Kubiznak}
 D. Kubiznak and R. B. Mann,
 {\em P-V criticality of charged AdS black holes},
  JHEP \textbf{1207}, 033 (2012),
   [arXiv:1205.0559 [hep-th]].

\bibitem{Weitwp}
 S.-W. Wei and Y.-X. Liu,
 {\em Triple points and phase diagrams in the extended phase space of charged Gauss-Bonnet black holes in AdS space},
  Phys. Rev. D \textbf{90}, 044057 (2014),
   [arXiv:1402.2837 [hep-th]].

\bibitem{Weicr}
 S.-W. Wei and Y.-X. Liu,
 {\em Thermodynamic nature of black holes in coexistence region},
  Sci. China Phys. Mech. Astron. \textbf{67}, 250412 (2024),
   [arXiv:2308.11886 [gr-qc]].

\end{thebibliography}
\end{document}